\documentclass[11pt]{article}       
\usepackage{geometry}               
\usepackage{graphicx}
\usepackage{caption}
\usepackage{subcaption}
\usepackage{enumerate}
\usepackage{hyperref}
\usepackage{upgreek}
\usepackage{array}
\usepackage{multirow}
\usepackage{mathtools}

\usepackage{amsmath} 
\usepackage{amssymb}  
\usepackage{amsthm}  
\usepackage{bm}
\usepackage{lipsum}
\usepackage[linesnumbered, ruled]{algorithm2e}
\usepackage{color}
\usepackage{cite}
\usepackage{tabularx}
\usepackage{makecell}

\newcolumntype{M}[1]{>{\centering\arraybackslash}m{#1}}

\newtheorem{proposition}{Proposition}

\newtheorem{assumption}{Assumption}

\newtheorem{lemma}{Lemma}
\newtheorem{theorem}{Theorem}
\newtheorem{remark}{Remark}

\newtheorem{cor}{Corollary}
\newcommand{\xpost}{\bar{x}}
\newcommand{\xprior}{\bar{x}^-}

\newcommand{\Xpost}{\Sigma_{x,t}}

\newcommand{\Xprior}{\Sigma_{x,t}^-}
\newcommand{\Xpriorhat}{\hat{\Sigma}_{x,t}^-}
\newcommand{\Xpostss}{\Sigma_{x,ss}}
\newcommand{\Xpriorss}{\Sigma_{x,ss}^-}
\newcommand{\Xpriorssnom}{\hat{\Sigma}_{x,ss}^{-}}

\newcommand{\Xpostssopt}{\Sigma_{x,ss}^*}
\newcommand{\Xpriorssopt}{\Sigma_{x,ss}^{-,*}}
\newcommand{\Xprioropt}{\Sigma_{x,t}^{-,*}}
\newcommand{\Xpostinit}{\Sigma_{x,0}}

\newcommand{\Xpriornom}{\hat{\Sigma}_{x,t}^-}
\newcommand{\Xpriornomnext}{\hat{\Sigma}^-_{x,t+1}}
\newcommand{\Xpriornomd}{\hat{\Sigma}_{x,t}^{-,d}}
\newcommand{\Xpriornomnextd}{\hat{\Sigma}_{x,t+1}^{-,d}}

\newcommand{\Xpriorinitopt}{\Sigma_{x,0}^{-,*}}

\newcommand{\Xpriornextopt}{\Sigma_{x,t+1}^{-,*}}
\newcommand{\Xpostprev}{\Sigma_{x,t-1}}
\newcommand{\xnom}{\hat{x}_0^-}
\newcommand{\Xnom}{\hat{\Sigma}_{x,0}^-}
\newcommand{\Pdist}{\mathbb{P}}
\newcommand{\Qdist}{\mathbb{Q}}
\newcommand{\Qhat}{\hat{\mathbb{Q}}}
\newcommand{\ambset}{\mathbb{D}}
\newcommand{\Gauss}{\mathcal{N}}

\newcommand{\Postdistopt}{\Pdist_{x,t}^*}
\newcommand{\Priordistopt}{\Pdist_{x,t}^{-,*}}
\newcommand{\Priordistinitopt}{\Pdist_{x,0}^{-,*}}

\newcommand{\Wass}{W_2}
\newcommand{\Bures}{\mathcal{B}}
\newcommand{\real}[1]{\mathbb{R}^{#1}}
\newcommand{\symm}[1]{\mathbb{S}^{#1}}
\newcommand{\psd}[1]{\symm{#1}_{+}}
\newcommand{\pd}[1]{\symm{#1}_{++}}

\DeclareMathOperator{\Tr}{Tr}

\usepackage{placeins}
\usepackage{cleveref}

\crefname{assumption}{assumption}{assumptions}

\title{On the Steady-State Distributionally Robust Kalman Filter\thanks{The first two authors contributed equally. This work was supported in part by the Information and Communications Technology Planning and Evaluation
(IITP) grants funded by MSIT No. 2022-0-00124, No. 2022-0-00480 and No. RS-2021-II211343,
Artificial Intelligence Graduate School Program (Seoul National University).}}

\author{Minhyuk Jang\thanks{Department of Mechanical Engineering, Seoul National University, Seoul, South Korea {\tt\small jason4012@snu.ac.kr}}
\and Astghik Hakobyan\thanks{Center for Scientific Innovation and Education and National Polytechnic University of Armenia, Yerevan, Armenia {\tt\small astghikhakobyan@csie.am}}
 \and
 Insoon Yang\thanks{Department of Electrical and Computer Engineering and ASRI, Seoul National University, Seoul, South Korea {\tt\small insoonyang@snu.ac.kr}}
}

\date{}

\begin{document}

\maketitle


\begin{abstract}
State estimation in the presence of uncertain or data-driven noise distributions remains a critical challenge in control and robotics. Although the Kalman filter is the most popular choice, its performance degrades significantly when distributional mismatches occur, potentially leading to instability or divergence. To address this limitation, we introduce a novel steady-state distributionally robust (DR) Kalman filter that leverages Wasserstein ambiguity sets to explicitly account for uncertainties in both process and measurement noise distributions. Our filter achieves computational efficiency by requiring merely the offline solution of a single convex semidefinite program, which yields a constant DR Kalman gain for robust state estimation under distributional mismatches. Additionally, we derive explicit theoretical conditions on the ambiguity set radius that ensure the asymptotic convergence of the time-varying DR Kalman filter to the proposed steady-state solution. Numerical simulations demonstrate that our approach outperforms existing baseline filters in terms of robustness and accuracy across both Gaussian and non-Gaussian uncertainty scenarios, highlighting its significant potential for real-world control and estimation applications.
\end{abstract}


\section{Introduction}\label{sec:intro}

State estimation is central to control and robotics, enabling systems to infer internal states from noisy and partial measurements. The Kalman filter~\cite{Kalman1960} remains a popular choice due to its simplicity and optimality under linear Gaussian assumptions. However, in practice, the true noise distributions are rarely known and are often estimated from limited data. As a result, applying the Kalman filter under such inaccurate models can lead to poor estimation accuracy and degraded control performance. This issue becomes even more critical in long-term operation, where both convergence and steady-state performance can no longer be guaranteed.

To address these challenges, robust filtering frameworks such as the $H_\infty$ filter~\cite{hassibi1999indefinite, simon2006optimal} have been proposed. While effective in bounding the worst-case energy gain from disturbances, these methods can be overly conservative in stochastic settings due to their lack of distributional modeling. In contrast, risk-sensitive filters~\cite{whittle1981risk, speyer1992optimal, hassibi2002linear} offer a probabilistic approach by penalizing large estimation errors via exponential cost functions. These filters maintain a Kalman-like structure but rely on a tunable robustness parameter that lacks a clear connection to the underlying model uncertainty.

In recent years, distributionally robust (DR) state estimation has emerged as a powerful framework for handling ambiguities in the noise distributions. Rather than relying on exact knowledge of the process and measurement noise distributions, DR state estimation seeks state estimators that perform well under the worst-case distributions within a pre-specified ambiguity set centered around a nominal model. Common choices for these ambiguity sets include divergence-based, moment-based, and distance-based sets~\cite{levy2012robust, wang2021distributionally, wang2022distributionally, zorzi2016robust, NEURIPS_DRKF, han2024distributionally}. 
A particularly appealing class of ambiguity sets is defined using the Wasserstein distance, which has received growing attention in both the optimization and control communities~\cite{esfahani2015data, kuhn2019wasserstein, gao2024wasserstein, blanchet2021statistical, yang2020wasserstein, hakobyan2024wasserstein, kim2023distributional, taskesen2024distributionally}. The Wasserstein metric provides strong geometric interpretability and is sensitive to the support of probability distributions, allowing it to meaningfully capture distributional shifts. 

In the context of robust filtering,\cite{NEURIPS_DRKF} proposed a Wasserstein-based DR Kalman filter using ambiguity sets centered around empirical distributions, leading to a tractable convex formulation with improved robustness against distributional mismatches. Building on this,\cite{han2024distributionally} proposed a DR Kalman filter that explicitly models covariance uncertainty by decoupling the ambiguity sets for process and measurement noises using a bicausal Wasserstein distance. Recent extensions include~\cite{lotidis2023wasserstein}, which incorporates martingale constraints to capture temporal dependence. In our prior work~\cite{jang2024wasserstein}, we developed a DR state estimator by recursively applying the DR minimum mean square error (DR-MMSE) estimator~\cite{nguyen2023bridging} with Wasserstein ambiguity sets, as part of a broader DR control framework for partially observable linear systems.

Despite recent progress in DR state estimation, most existing DR Kalman filtering methods are formulated in a time-varying fashion, which is computationally expensive---especially in long-horizon or real-time applications. By contrast, under certain conditions, the standard Kalman filter is known to converge to a steady-state solution with a constant gain, offering both computational efficiency and stable long-term performance. Similar convergence results have been established for risk-sensitive and robust Kalman filters under specific conditions on the robustness parameters~\cite{zorzi2015convergence, zorzi2017convergence}.

Motivated by this, we propose a time-invariant formulation of the Wasserstein DR Kalman filter under stationary Gaussian nominal distributions. Our approach explicitly incorporates distributional uncertainty in both process and measurement noises via Wasserstein ambiguity sets. Crucially, our DR Kalman filter requires solving a single semidefinite program (SDP) \emph{offline} to determine the least-favorable covariance matrices and the corresponding constant DR Kalman gain. This differs from~\cite{kargin2024distributionally} and~\cite{hajar2025distributionally}, which analyze the infinite-horizon performance of the DR filter in terms of cumulative estimation error without explicitly accounting for the recursive structure of state estimation.

We further provide theoretical guarantees for the convergence of the time-varying DR Kalman filter to the proposed time-invariant filter by analyzing the DR Riccati recursion.  In particular, we derive explicit conditions on the ambiguity set radius that ensure the DR Riccati mapping becomes a contraction. Leveraging these conditions, we show that the time-varying DR Kalman filter asymptotically converges to its steady-state counterpart, matching our proposed time-invariant filter. Owing to its structural similarity with the steady-state Kalman filter, our proposed method remains straightforward to implement and computationally efficient, making it suitable for practical applications. At the same time, it effectively mitigates distributional mismatches by accounting for worst-case deviations from nominal distributions. Finally, our simulations confirm that the DR Kalman filter converges to the steady-state solution with the derived ambiguity radius and consistently outperforms baselines under both Gaussian and non-Gaussian uncertainties.

The remainder of the paper is organized as follows. In~\Cref{sec:prel}, we present the problem setup and introduce the DR state estimation framework using Wasserstein ambiguity sets. In~\Cref{sec:DRKF}, we review the time-varying DR Kalman filter and its tractable SDP implementation. In~\Cref{sec:infDRKF}, we propose the time-invariant DR Kalman filter and analyze the convergence of the time-varying filter to its steady-state counterpart. Finally,~\Cref{sec:experi} evaluates the proposed method through numerical experiments demonstrating both convergence and control performance.

\section{Preliminaries}\label{sec:prel}

\subsection{Notation}

Let $\mathbb{S}^n$ denote the space of all symmetric matrices in $\mathbb{R}^{n\times n}$. We use $\mathbb{S}_+^n$ (respectively, $\mathbb{S}_{++}^n$) to represent the cone of symmetric positive semidefinite (positive definite) matrices in $\mathbb{S}^n$. 
For any $A,B\in\mathbb{S}^n$, the relation $A\succeq B$ (respectively, $A \succ B$) means that  $A - B\in\mathbb{S}_+^n$ (respectively, $A - B \in \mathbb{S}_{++}^n$). 
The set of Borel probability measures with support $\mathcal{W}\subseteq \mathbb{R}^n$ is denoted by $\mathcal{P}(\mathcal{W})$, while the subset of measures with finite second-order moments is denoted by $\mathcal{P}_2(\mathcal{W})$. For a matrix $A \in \mathbb{S}^n$, we denote its $i$-th eigenvalue by $\lambda_i(A)$, and its smallest and largest eigenvalues by $\lambda_{\min}(A)$ and $\lambda_{\max}(A)$, respectively. For a sequence of vectors $w_k \in \mathbb{R}^n$, we define the stacked vector for $k = tN, \dots, tN + N - 1$ as $\bm{w}_t^N \coloneqq
    [w_{tN+N-1}^\top, \dots, w_{tN}^\top]^\top$.

\subsection{Problem Setup}\label{sec:problem_setup}

Consider the following discrete-time linear stochastic system:  
\begin{equation}\label{eqn:system_eq}
\begin{split}
    x_{t+1} &= A x_{t} + w_{t},\\
     y_t &= C x_t + v_t,
\end{split}
\end{equation}
where $x_t\in\real{n_x}$ and $y_t \in\real{n_y}$ denote the system state and measurement at time $t$, respectively. The process noise $w_t \in \real{n_x}$ and the measurement noise $v_t \in \real{n_y}$ are governed by probability distributions $\Qdist_{w,t} \in \mathcal{P}(\mathbb{R}^{n_x})$ and $\Qdist_{v,t} \in \mathcal{P}(\real{n_y})$, respectively. The initial state $x_0$ is sampled from an initial distribution $\mathbb{Q}_{x,0} \in \mathcal{P}(\mathbb{R}^{n_x})$. Additionally, we assume that $x_0$, $\{w_t\}$, and $\{v_t\}$ are mutually independent, and that $w_t$ and $w_{t'}$ (similarly, $v_t$ and $v_{t'}$) are independent for $t \neq t'$.

At each time $t$, we define the accumulated measurement history as $\mathcal{Y}_t \coloneqq \{y_0, \dots, y_t\}, \quad t \geq 0$.
Our objective is to estimate the state $x_t$ based on the available measurements $\mathcal{Y}_t$. Since measurements are sequentially observed, a natural approach is to recursively compute the minimum mean square error (MMSE) estimator:
\begin{equation}\label{eqn:MMSE}
    \min_{\phi_t \in \mathcal{F}_t} J_{MSE}(\phi_t, \Qdist_t)  \coloneqq \mathbb{E}_{\Pdist_t}[\|x_t - \phi_t(y_t)\|^2 \mid \mathcal{Y}_{t-1}],
\end{equation}
where $\mathcal{F}_t = \{ \phi_t \mid \phi_t(y_t) \in \real{n_x},  \phi_t \text{ is measurable} \}$ denotes the set of admissible estimators. Here, $\Pdist_t$ represents the joint distribution of $x_t$ and $y_t$, which, under model~\eqref{eqn:system_eq}, factorizes as $\Qdist_t = \Qdist_{x,t}^- \otimes \Qdist_{v,t}$, with $\Qdist_{x,t}^-$ being the prior distribution of $x_t$ given the past measurements $\mathcal{Y}_{t-1}$.

In the classical setting, when $x_0$, $\{w_t\}$, and $\{v_t\}$ are zero-mean Gaussian with known covariance matrices, the Kalman filter provides the optimal MMSE estimator~\eqref{eqn:MMSE}. In practice, however, the noise statistics are typically estimated from limited data and are therefore subject to ambiguity. Specifically, only nominal estimates $\hat{\Qdist}_{w,t}, \hat{\Qdist}_{v,t},$ and $\hat{\Qdist}_{x,0}^-$ of the noise distributions are available. Applying a Kalman filter based on these inaccurate models can significantly degrade performance and may even cause divergence. This sensitivity is especially problematic in the steady-state regime. While the classical Kalman filter is guaranteed to converge under certain conditions, such convergence and long-term performance cannot be assured when the underlying distributions are uncertain.
To address this challenge, we propose a time-invariant DR state estimation framework that explicitly accounts for ambiguities in the noise distributions.

\subsection{DR State Estimation with Wasserstein Ambiguity Sets}

The goal of the DR state estimation framework is to design an optimal state estimator that remains robust against inaccuracies in the nominal uncertainty distributions. To achieve this, we reformulate the classical MMSE estimation problem~\eqref{eqn:MMSE} as the following DR-MMSE estimation problem:
\begin{equation}\label{eqn:DRMMSE}
\min_{\phi_t \in \mathcal{F}_t} \max_{\Pdist_t \in \ambset_\theta(\hat{\Qdist}_t)} J_{MSE}(\phi_t, \Pdist_t),
\end{equation}
where the inner maximization selects the least-favorable joint prior state and measurement noise distributions from the ambiguity set $\ambset_\theta(\hat{\Qdist}_t)$, which is centered around the nominal distribution $\hat{\Qdist}_t \coloneqq \hat{\Qdist}_{x,t}^- \otimes \hat{\Qdist}_{v,t}$, representing the data-driven estimate of $\Qdist_t$.

In this work, we construct the ambiguity set using the Wasserstein distance, which provides a notion of distance between probability distributions. Specifically, we define the Wasserstein ambiguity set as 
\begin{equation*}\label{eqn:amb_set}
\begin{split}
    \ambset_\theta(\hat{\Qdist}_t) = & \Big\{ \Pdist_{x,t}^- \otimes \Pdist_{v,t} \mid \Pdist_{x,t}^- \in \mathcal{P}_2(\real{n_x}), \Pdist_{v,t} \in \mathcal{P}_2(\real{n_y}),\\
    & \Wass(\Pdist_{x,t}^-, \hat{\Qdist}_{x,t}^-) \leq \theta_x, \Wass(\Pdist_{v,t}, \hat{\Qdist}_{v,t}) \leq \theta_v \Big\},
\end{split}
\end{equation*}
where $\Wass$ denotes the 2-Wasserstein distance, measuring the discrepancy between two probability distributions in terms of the optimal transport cost. Specifically, for two distributions $\Pdist_1, \Pdist_2 \in \mathcal{P}(\mathcal{W})$, the 2-Wasserstein distance is given by 
\begin{align*}
    \Wass(\Pdist_1, \Pdist_2) \coloneqq \inf_{\tau \in \mathcal{T}(\Pdist_1, \Pdist_2)} \left\{ \left(\int_{\mathcal{W} \times \mathcal{W}} {\|x - y\|}^2 d\tau(x, y) \right)^{\frac{1}{2}} \right\},
\end{align*}
where $\| \cdot \|$ is an arbitrary norm defined on $\mathcal{W}$, and $\mathcal{T}(\Pdist_1, \Pdist_2)$ is the set of joint probability distributions on $\mathcal{W} \times \mathcal{W}$ with marginals $\Pdist_2$ and $\Pdist_2$, respectively.\footnote{In this work, we adopt the standard Euclidean norm for $\|\cdot\|$.} 

The Wasserstein ambiguity set offers several attractive properties, such as avoiding degenerate or pathological distributions and providing a meaningful notion of proximity between probability measures based on the underlying geometry of their supports. As a result, it serves as a powerful foundation for developing DR versions of classical estimators, including the Kalman filter.

However, directly solving the minimax problem~\eqref{eqn:DRMMSE} under Wasserstein ambiguity is generally intractable, as it involves optimization over an infinite-dimensional space of probability measures. Fortunately, a key result from~\cite{gelbrich1990formula} provides an elegant simplification when the involved distributions are Gaussian. Specifically, if $\Pdist_1=\Gauss(\mu_1,\Sigma_1)$ and $\Pdist_2=\Gauss(\mu_2,\Sigma_2)$ are Gaussians with mean vectors  $\mu_1, \mu_2 \in \real{n}$ and covariance matrices $\Sigma_1, \Sigma_2 \in \psd{n}$, then the 2-Wasserstein distance admits the closed-form expression:
\begin{equation}\label{eqn:gelb}
\Wass(\Pdist_1, \Pdist_2) = \sqrt{\|\mu_1 - \mu_2\|_2^2 + \Bures^2(\Sigma_1, \Sigma_2)},
\end{equation}
where
\[
\Bures(\Sigma_1, \Sigma_2) \coloneqq \left(\Tr\left[\Sigma_1 + \Sigma_2 - 2\left(\Sigma_2^{\frac{1}{2}} \Sigma_1 \Sigma_2^{\frac{1}{2}}\right)^{\frac{1}{2}}\right]\right)^{\frac{1}{2}}
\]
denotes the Bures-Wasserstein distance between covariance matrices.
While this formula is exact for Gaussian distributions, it also provides a tractable and meaningful lower bound on the Wasserstein distance between arbitrary distributions with finite second-order moments. By relying only on the first- and second-order moments, the expression~\eqref{eqn:gelb} enables the reformulation of the original intractable DR state estimation problem into a finite-dimensional convex optimization problem.

\section{Distributionally Robust Kalman Filter}\label{sec:DRKF}

As discussed in~\Cref{sec:problem_setup}, the standard Kalman filter recursively solves the MMSE estimation problem~\eqref{eqn:MMSE} under the assumption of zero-mean Gaussian noise. In parallel, this section introduces a tractable framework for DR state estimation in the form of a DR Kalman Filter, which iteratively solves the DR-MMSE estimation problem~\eqref{eqn:DRMMSE} under nonzero-mean Gaussian nominal distributions while ensuring robustness against distributional uncertainties.

\begin{assumption}
\label{ass:Gauss}
The nominal distributions for the initial state $x_0$, process noise $w_t$, and measurement noise $v_t$ for all $t$ are Gaussian, i.e., $\Qhat_{x,0} = \mathcal{N}(\xnom, \Xnom), 
\Qhat_{w,t} = \mathcal{N}(\hat{w}_t,\hat{\Sigma}_{w,t}), 
\Qhat_{v,t} = \mathcal{N}(\hat{v}_t,\hat{\Sigma}_{v,t})$,
where $\hat{x}_0, \hat{w}_t\in\real{n_x}$ and $\hat{v}_t\in\real{n_y}$ are the mean vectors, while $\Xnom, \hat{\Sigma}_{w,t} \in \psd{n_x}$ and $\hat{\Sigma}_{v,t}\in\real{n_y}$ are the covariance matrices.
\end{assumption}

Notably, we impose no assumptions on the actual uncertainty distributions, allowing them to be non-Gaussian. Given~\Cref{ass:Gauss}, we reformulate the DR-MMSE estimation problem into a finite convex optimization problem.

\begin{lemma}[Wasserstein DR-MMSE Estimation Problem {\cite[Theorem 3.1]{nguyen2023bridging}}]
\label{lem:GelbrichMMSE}
Suppose~\Cref{ass:Gauss} holds. Then, at any time stage $t\geq 0$, the DR-MMSE estimation problem~\eqref{eqn:DRMMSE} given the nominal distributions $\hat{\Qdist}_{x,t}^- = \mathcal{N}(\bar{x}_t^-, \Xpriornom)$ and $\hat{\Qdist}_{v,t} = \mathcal{N}(\hat{v}_t, \hat{\Sigma}_{v,t})$ is equivalent to the following finite convex optimization problem:
\begin{equation}\label{eqn:DRMMSE_opt}
\begin{split}
\max_{\Sigma_{v,t}, \Xprior} \; & \Tr[\Xprior-\Xprior C^{\top}  (C\Xprior C^{\top} + \Sigma_{v,t})^{-1} C \Xprior] \\
\mbox{s.t.} \; & \Tr[\Sigma_{v,t} + \hat{\Sigma}_{v,t} -2 \big(\hat{\Sigma}_{v,t}^\frac{1}{2} \Sigma_{v,t} \hat{\Sigma}_{v,t}^\frac{1}{2}\big)^{\frac{1}{2}}] \leq \theta_{v}^2\\
& \Tr[\Xprior + \Xpriornom -2 \big((\Xpriornom)^\frac{1}{2} \Xprior (\Xpriornom)^\frac{1}{2}\big)^{\frac{1}{2}}] \leq \theta_{x}^2\\
& \Sigma_{v,t}\in\pd{n_y}, \Xprior\in\psd{n_x}.
\end{split}
\end{equation}

Moreover, if $(\Xprioropt,\Sigma_{v,t}^{*})$ is an optimal solution of~\eqref{eqn:DRMMSE_opt}, then the maximum in~\eqref{eqn:DRMMSE} is attained by the Gaussian distributions $\Priordistopt = \Gauss(\bar{x}_t^-, \Xprioropt)$ and $\Pdist_{v,t}^{*} = \Gauss(\hat{v}_t, \Sigma_{v,t}^{*})$ for each $t\geq 0$. 
Furthermore, at each time step $t$, the optimal estimator attaining the minimum in~\eqref{eqn:DRMMSE} is given by:
\begin{equation}\label{eqn:DR_estimator}
\begin{split}
\phi_{t}^*(y_{t}) = \xprior_t +  \Xprioropt C^{\top} (C\Xprioropt & C^{\top}+ \Sigma_{v,t}^{*})^{-1} \times (y_{t}-C\xprior_t -\hat{v}_{t}),
\end{split}
\end{equation}
with $\xprior_0 = \xnom$.
\end{lemma}

Note that the problem~\eqref{eqn:DRMMSE_opt} (with $ \hat{\Sigma}_{v,t} \in \pd{n_y}$) is solvable as it has a continuous objective function over a compact feasible set.
With the finite reformulation of the DR-MMSE estimation problem in hand, the following theorem demonstrates how it can be iteratively solved, leading to the DR version of the Kalman filter.

\begin{theorem}[DR Kalman Filter]\label{thm:DRKF}
    Suppose~\Cref{ass:Gauss} holds. Then, for the system~\eqref{eqn:system_eq} with the corresponding nominal distributions, the DR state estimates $\phi^*_t(y_t)$, which recursively solve~\eqref{eqn:DRMMSE}, coincide with the conditional expectation $\xpost_t$ of the states. 
    Additionally, for each time stage $t$, the least-favorable prior and posterior state distributions retain Gaussian forms: $\Priordistopt = \Gauss(\xprior_t, \Xprioropt)$ and $\Postdistopt = \Gauss(\xpost_t, \Xpost)$ with $\xprior_0 = \xnom$. These are computed recursively as follows:
\begin{itemize}
\item \textbf{(Measurement Update)} Update $\xpost_t$ and $\Xpost$ based on the measurement $y_t$ as follows:
\begin{align}
\xpost_t &= \xprior_t + K_t^* (y_{t}-C \xprior_t -\hat{v}_{t}) \label{cond_mean}\\
\Xpost &= (I-K_t^*  C) \Xprioropt,\label{cond_cov}
\end{align}
where
\begin{equation}\label{eqn:Kalman_gain}
K_t^* = \Xprioropt C^{\top} (C \Xprioropt C^{\top} + \Sigma_{v,t}^{*})^{-1}
\end{equation}
is the DR Kalman gain matrix, and $(\Xprioropt, \Sigma_{v,t}^{*})$ is a maximizing pair of~\eqref{eqn:DRMMSE_opt} for all $t$, given $\Xpriornom$ and $\hat{\Sigma}_{v,t}$.

\item \textbf{(State Prediction)} Predict $\xprior_{t+1}$ and the \emph{pseudo-nominal} prior state covariance matrix $\Xpriornomnext$ as follows:
\begin{align}
    \begin{split}
        \xprior_{t+1} &= A \xpost_{t} + \hat{w}_t \label{cond_mean_prior}
    \end{split} \\
    \Xpriornomnext &= A \Xpost A^\top + \hat{\Sigma}_{w,t}\label{cond_cov_prior}
\end{align}
\end{itemize}

\end{theorem}

The proof of this theorem can be found in Appendix~\ref{app:DRKF}.
The measurement update and state prediction equations~\eqref{cond_mean}--\eqref{cond_cov_prior} closely follow those of the standard Kalman filter. However, in the state prediction step, the DR Kalman filter introduces the pseudo-nominal mean vector and covariance matrix, to account for distributional uncertainty. Additionally, due to the nonzero mean assumption, extra terms appear in both the prior and posterior state mean computations, capturing the influence of the nominal mean vectors of the process and measurement noise.

\begin{remark}
The proposed DR Kalman filter returns the optimal DR-MMSE estimator as long as~\Cref{ass:Gauss} is satisfied, even when the actual uncertainty distributions are non-Gaussian. Furthermore, similar to the standard Kalman filter, if the nominal distributions themselves deviate from Gaussianity, the DR Kalman filter remains applicable and yields the best DR linear unbiased state estimator.
\end{remark}

It has been shown in~\cite{hakobyan2024wasserstein} that the DR-MMSE estimation problem of the form~\eqref{eqn:DRMMSE_opt} can be equivalently reformulated into a tractable SDP problem. Specifically, by employing the standard Schur complement argument (e.g.,~\cite[Appendix 5.5]{boyd2004convex}), it is equivalent to the following optimization problem:
\begin{equation} \label{eqn:DRKF_SDP}
\begin{split}
\max_{\substack{\Xprior, \Xpost\\ \Sigma_{v,t},Y,Z \\ }} \; & \Tr[\Xpost ] \\
\mbox{s.t.} \; & \begin{bmatrix} \Xprior - \Xpost & \Xprior C^{\top} \\ C \Xprior & C \Xprior C^{\top} + \Sigma_{v,t} \end{bmatrix} \succeq 0\\
& \begin{bmatrix} \Xpriornom & Y \\ Y^{\top} & \Xprior \end{bmatrix} \succeq 0, \quad \begin{bmatrix} \hat{\Sigma}_{v,t} & Z \\ Z^{\top} & \Sigma_{v,t} \end{bmatrix} \succeq 0 \\
&\Tr[\Xprior + \Xpriornom  -2 Y]  \leq \theta_x^2 \\
&\Tr[\Sigma_{v,t} + \hat{\Sigma}_{v,t} -2 Z] \leq \theta_v^2\\
& \Xprior, \Xpost \in \psd{n_x}, \; \Sigma_{v,t} \in \pd{n_y}\\
&Y\in\real{n_x\times n_x},\; Z \in\real{n_y \times n_y}.
\end{split}
\end{equation}
Hence, solving this SDP in each time step provides least-favorable covariance matrices $(\Xprioropt, \Sigma_{v,t}^{*})$, thereby making the estimator robust against distributional ambiguities. 

\section{Steady-State Distributionally Robust Kalman Filter}\label{sec:infDRKF}

Although the DR Kalman filter is computationally tractable and provides DR state estimates, it requires solving the SDP problem~\eqref{eqn:DRKF_SDP} at every measurement update step. While this is feasible,  the associated computational burden poses challenges for real-time applications. To address this issue, we propose a time-invariant DR Kalman filtering algorithm in which a single SDP problem is solved offline to determine a constant DR Kalman gain matrix. We further establish that, under certain conditions, the DR Kalman filter asymptotically converges to its steady-state counterpart, corresponding to our time-invariant filter.

\subsection{Time-Invariant DR Kalman Filtering Algorithm}

To derive the time-invariant version of the DR Kalman filter, we introduce the following assumption.
\begin{assumption}\label{ass:time_inv_nom}
    The nominal state distributions are stationary, and thus $\hat{w}_t = \hat{w}, \hat{v}_t=\hat{v}$ and $\hat{\Sigma}_{v,t}=\hat{\Sigma}_v, \hat{\Sigma}_{w,t}=\hat{\Sigma}_w$ for all $t$. Moreover, the covariance matrices are positive definite, i.e., $\hat{\Sigma}_w \in \pd{n_x}$ and $\hat{\Sigma}_v \in \pd{n_y}$.
\end{assumption}

Since the nominal distributions remain unchanged over time, instead of solving the time-varying optimization problem~\eqref{eqn:DRMMSE} (or equivalently the SDP~\eqref{eqn:DRKF_SDP}) at every time step, we propose the following time-invariant optimization problem:
\begin{equation}\label{eqn:ss_DRKF}
    \begin{split}
\max_{\substack{\Xpriorssnom ,\Xpriorss\\ \Xpostss , \Sigma_{v,ss}}} \; & \Tr[\Xpostss ] \\
\mbox{s.t.} \; & \Xpostss = \Xpriorss-\Xpriorss C^{\top} (C\Xpriorss C^{\top} + \Sigma_{v,ss})^{-1} C \Xpriorss\\
& \Xpriorssnom = A \Xpostss A^\top + \hat{\Sigma}_{w}\\
&\Tr[\Xpriorss + \Xpriorssnom  -2 \big((\Xpriorssnom)^\frac{1}{2} \Xpriorss (\Xpriorssnom)^\frac{1}{2})^{\frac{1}{2}}]  \leq \theta_x^2 \\
&\Tr[\Sigma_{v,ss} + \hat{\Sigma}_{v} -2 \big(\hat{\Sigma}_{v}^\frac{1}{2} \Sigma_{v,ss} \hat{\Sigma}_{v}^\frac{1}{2}\big)^{\frac{1}{2}}] \leq \theta_v^2\\
& \Xpriorssnom ,\Xpriorss, \Xpostss \in \psd{n_x}, \; \Sigma_{v,ss} \in \pd{n_y}.
\end{split}
\end{equation}
The optimization problem~\eqref{eqn:ss_DRKF} effectively combines the covariance update equations from both the measurement update and state prediction phases. Consequently, it is sufficient to solve this problem once offline to obtain the least-favorable prior and posterior state covariance matrices, denoted by $\Xpriorssopt$ and $\Xpostssopt$, along with the least-favorable measurement noise covariance matrix $\Sigma_{v,ss}^*$.

\begin{algorithm}[t]
\DontPrintSemicolon
\SetKw{Input}{Input:}
\SetKw{to}{to}
\SetKw{and}{and}
\SetKw{break}{break}
\Input $\theta_x,  \theta_{v} , \hat{\Sigma}_w, \hat{\Sigma}_v, \xprior_0$\;
{Solve~\eqref{eqn:ss_DRKF_SDP} offline to obtain $(\Xpriorssopt$, $\Xpostssopt, \Sigma_{v,ss}^{*})$}\;
Compute the DR Kalman gain $K_{ss}^* = \Xpriorssopt C^{\top} (C\Xpriorssopt  C^{\top}+ \Sigma_{v,ss}^*)^{-1}$\;
\For{$t=0,1\ldots$}{
Observe $y_{t}$\;
Update $\xpost_t = \xprior_t +  K_{ss}^* (y_{t}-C\xprior_t -\hat{v})$\;
Estimate $\xprior_{t+1} = A\xprior + \hat{w}$\;
}
\caption{Time-Invariant DR Kalman Filter} \label{alg:inf_DRKF}
\end{algorithm}

Similar to~\eqref{eqn:DRKF_SDP}, we can apply the Schur complement argument to reformulate~\eqref{eqn:ss_DRKF} into the following tractable SDP problem:
\begin{equation} \label{eqn:ss_DRKF_SDP}
\begin{split}
\max_{\substack{\Xpriorssnom ,\Xpriorss,\\ \Xpostss, \Sigma_{v,ss},\\ Y,Z}} \; & \Tr[\Xpostss ] \\
\mbox{s.t.} \; & \begin{bmatrix} \Xpriorss - \Xpostss & \Xpriorss C^{\top} \\ C \Xpriorss & C \Xpriorss C^{\top} + \Sigma_{v,ss} \end{bmatrix} \succeq 0\\
& \begin{bmatrix} \Xpriorssnom & Y \\ Y^{\top} & \Xpriorss \end{bmatrix} \succeq 0, \quad \begin{bmatrix} \hat{\Sigma}_{v} & Z \\ Z^{\top} & \Sigma_{v,ss} \end{bmatrix} \succeq 0 \\
& \Xpriorssnom = A \Xpostss A^{\top} + \hat{\Sigma}_{w} \\
&\Tr[\Xpriorss + \Xpriorssnom  -2 Y]  \leq \theta_x^2 \\
&\Tr[\Sigma_{v,ss} + \hat{\Sigma}_{v} -2 Z] \leq \theta_v^2\\
& \Xpriorssnom ,\Xpriorss, \Xpostss \in \psd{n_x}, \; \Sigma_{v,ss} \in \pd{n_y}\\
&Y\in\real{n_x\times n_x},\; Z \in\real{n_y \times n_y}.
\end{split}
\end{equation}
Let $(\Xpriorssopt, \Xpostssopt, \Sigma_{v,ss}^*)$ be an optimal solution to~\eqref{eqn:ss_DRKF_SDP}. Then, the time-invariant DR state estimator is defined as
\begin{equation*} \label{eqn:DR_est_update} 
\phi_{ss}^*(y_{t}) := \xprior_t +  K_{ss}^* (y_{t}-C\xprior_t -\hat{v}), 
\end{equation*}
where the constant DR Kalman gain matrix is given by
\[
K_{ss}^* := \Xpriorssopt C^{\top} (C\Xpriorssopt  C^{\top}+ \Sigma_{v,ss}^*)^{-1}.
\]

The overall time-invariant DR Kalman filtering algorithm is described in~\Cref{alg:inf_DRKF}. In the offline stage, the least-favorable matrices $(\Xpriorssopt,\, \Xpostssopt,\, \Sigma_{v,ss}^*)$ and the constant DR Kalman gain $K_{ss}^*$ are computed by solving~\eqref{eqn:ss_DRKF_SDP}. In the online stage, state estimates are updated upon receiving new observations. 
An important practical advantage of our time-invariant DR Kalman filter is that, once the least-favorable covariance matrices are obtained offline, the filter can be implemented identically to the standard time-invariant Kalman filter. Overall, the proposed filter offers significant advantages in practical applications---particularly when the estimation horizon is unknown or infinite---by being more scalable than the time-varying DR Kalman filter, which requires solving an optimization problem at every time stage.

\subsection{Convergence of the DR Kalman filter}

Having introduced both the time-varying and time-invariant DR Kalman filters, we now analyze the convergence of the time-varying filter to its steady-state counterpart. In particular, we seek to establish conditions under which the time-invariant filter derived above corresponds to the steady-state DR Kalman filter. That is, we aim to identify conditions on the Wasserstein ambiguity set under which there exists a constant DR Kalman gain matrix $K_{ss}$ satisfying
\[
K_{ss} = \lim_{t \to \infty} K_t^*.
\]

To guarantee the existence of such a steady-state solution, we impose the following standard assumptions:
\begin{assumption}\label{ass:ctrb_obsv} The pair $(A,\hat{\Sigma}_w^{\frac{1}{2}})$ is controllable and the pair $(A, C)$ is observable.
\end{assumption}

To analyze the convergence, consider the following Riccati-like mapping, referred to as the \emph{DR Riccati equation}:
\begin{equation}\label{eqn:riccati_like}
    r_\ambset(\Xpriornom) \coloneqq A((\Xprioropt)^{-1} + C^\top (\Sigma_{v,t}^*)^{-1} C)^{-1} A^\top + \hat{\Sigma}_{w},
\end{equation}
where $(\Xprioropt, \Sigma_{v,t}^{*})$ is an optimal solution pair to~\eqref{eqn:DRMMSE_opt} given the pseudo-prior covariance matrix $\Xpriornom$. The inverse of $\Xprioropt$ exists because any optimal solution to~\eqref{eqn:DRMMSE} satisfies the condition $\Xprioropt \succeq \lambda_{\min}(\Xpriornom) I_{n_x}$, which, combined with~\Cref{ass:time_inv_nom} implies that $\Xprioropt$ is positive definite.

By applying the matrix inversion lemma to~\eqref{eqn:riccati_like}, the pseudo-nominal prior state covariance matrix update equation~\eqref{cond_cov_prior} becomes
\begin{equation*}\label{eqn:woodbury}
    \Xpriornomnext = r_\ambset(\Xpriornom)
\end{equation*}
with initial condition $\Xnom \in \pd{n_x}$.
Define
\[
\Phi_t \coloneqq (\Xpriornomnext)^{-1} - (\Xpriornextopt)^{-1} + C^{\top} (\hat{\Sigma}_v^{-1} - (\Sigma_{v,t}^{*})^{-1}) C.
\]
Then, the DR Riccati equation~\eqref{eqn:riccati_like} can be rewritten as
\begin{equation}\label{eqn:DR_riccati}
r_\ambset(\Xpriornom) = A((\Xpriornom)^{-1} + C^{\top} \hat{\Sigma}_v^{-1} C - \Phi_{t-1}  )^{-1} A^\top + \hat{\Sigma}_{w}.
\end{equation}

Remarkably, the DR Riccati mapping~\eqref{eqn:DR_riccati} shares the same structural form as the well-known risk-sensitive Riccati equation~\cite{whittle1981risk, jacobson2003optimal}. In~\cite{levy2016contraction}, it was shown that under~\Cref{ass:ctrb_obsv}, the $N$-fold composition of the risk-sensitive Riccati mapping becomes a contraction with respect to the Riemannian metric on the manifold of positive definite matrices, provided that $N \geq n_x$. Extending these results,~\cite{zorzi2017convergence} demonstrated similar contraction properties for the Thompson part metric, applying to a broader class of robust Kalman filters. 
These observations lay the groundwork for establishing convergence of the DR Kalman filter by drawing parallels to the contraction behavior observed in risk-sensitive and robust filters.

As established in~\cite{zorzi2017convergence}, the Riccati-like mapping~\eqref{eqn:riccati_like}, due to its structural similarity to the risk-sensitive Riccati equation, admits an equivalent interpretation of the DR Kalman filter as solving a least-squares filtering problem with time-varying parameters in a Krein space. In our setting, we consider the system~\eqref{eqn:system_eq} under nominal stationary distributions and denote by $\hat{x}_t$ and $\hat{y}_t$ the corresponding nominal state and measurement vectors. By augmenting the original system with an auxiliary uncertainty vector $u_t \in \mathbb{R}^{n_x}$ and introducing a synthetic observation of the form $0 = \hat{x}_t + u_t$, the combined uncertainty vectors $w_t$, $v_t$, and $u_t$ associated with the nominal system are viewed as elements of a Krein space with the inner product
\[
\left\langle \begin{bmatrix}
    w_t \\ v_t \\ u_t
\end{bmatrix}, \begin{bmatrix}
    w_s \\ v_s \\ u_s
\end{bmatrix} \right\rangle = \begin{bmatrix}
    \hat{\Sigma}_{w} & 0 & 0\\ 0 & \hat{\Sigma}_v & 0 \\ 0 & 0 & -\Phi_t^{-1}
\end{bmatrix}\delta_{t-s},
\]
where $\delta_t$ denotes the Kronecker delta function.

To proceed with the convergence analysis, we introduce the downsampled nominal state vector $\hat{x}_t^d = \hat{x}_{tN}$ for some $N \in \mathbb{N}$. Since $\hat{x}_t$ is Gaussian under the nominal distribution, the downsampled states $\hat{x}_t^d$ remain Gaussian and evolve according to the following state-space model:
\[
\begin{split}
    \hat{x}_{t+1}^d &= A^N \hat{x}_t^d + \mathcal{R}_N \bm{w}_t^N\\
    \hat{y}_t^N &= \mathcal{O}_N \hat{x}_t^d + \mathcal{D}_N \bm{v}_t^N + \mathcal{H}_N \bm{w}_t^N\\
    \bm{0} &= \mathcal{O}_N^R \hat{x}_t^d + \bm{u}_t^{N} + \mathcal{L}_N \bm{w}_t^N,
\end{split}
\]
with 
\[
\begin{split}
\mathcal{R}_N &\coloneqq 
    \left[
        \hat{\Sigma}_w^{\frac{1}{2}}, \, A \hat{\Sigma}_w^{\frac{1}{2}}, \,\ldots,\, A^{N-1} \hat{\Sigma}_w^{\frac{1}{2}}
    \right]
    \\
    \mathcal{O}_N &\coloneqq
    \left[
        (CA^{N-1})^\top,\, \ldots, \,(CA)^\top,\, C^\top
    \right]^\top\\
    \mathcal{O}_N^R &\coloneqq 
\left[(A^{N-1})^\top,\, \ldots,\, (A)^\top,\, I_{n_x}\right]^\top\\
    \mathcal{D}_N & \coloneqq I_N \otimes \hat{\Sigma}_v^{\frac{1}{2}},
\end{split}
\]
where $\mathcal{R}_N$ and $\mathcal{O}_N$ represent the $N$-block controllability and observability matrices.
The block Henkel matrices $\mathcal{L}_N$ and $\mathcal{H}_N \coloneqq (I_N \otimes C) \mathcal{L}_N$ encode cross-correlations between observations and noise terms, with entries of $\mathcal{L}_N$ defined as $[\mathcal{L}_N]_{ij} = A^{j-i-1} \hat{\Sigma}_w^{\frac{1}{2}}$ for $j>i$ and zero otherwise.

Define the block diagonal matrix
\[
\bar{\Phi}_{N,t} := \mathrm{blkdiag}(\Phi_{tN+N-2}, \Phi_{tN+N-3},\dots, \Phi_{t N-1}).
\]

Following the procedure in~\cite{zorzi2017convergence} and  letting $\Xpriornomd := \hat{\Sigma}_{x,tN}^-$ denote the downsampled pseudo-nominal covariance matrix, the DR Riccati recursion for the downsampled model becomes
\[
\Xpriornomnextd = r_{\ambset,t}^d(\Xpriornomd),
\]
where
\begin{equation}\label{eqn:downsampled_Riccati}
    r_{\ambset,t}^d(\Xpriornomd) \coloneqq \alpha_{N,t}\left[ (\Xpriornomd)^{-1} + \Omega_{\bar{\Phi}_{N,t}} \right]^{-1} \alpha_{N,t} + W_{\bar{\Phi}_{N,t}},
\end{equation}
with
\begin{equation*}\label{eqn:ds_matrices}
\begin{split}
    \alpha_{N,t} &\coloneqq A^N - \mathcal{R}_N (\mathcal{H}_N^\top \mathcal{K}_{{\bar{\Phi}}_{N,t}}^{-1} \mathcal{O}_N + \mathcal{L}_N^\top  \mathcal{K}_{{\bar{\Phi}}_{N,t}}^{-1} \mathcal{O}_N^R)\\
    W_{\bar{\Phi}_{N,t}} &\coloneqq \mathcal{R}_N \mathcal{Q}_{\bar{\Phi}_{N,t}}\mathcal{R}_N^\top\\
    \mathcal{Q}_{\bar{\Phi}_{N,t}} & \coloneqq \left[ I_{N n_x} + \mathcal{H}_N^\top (\mathcal{D}_N \mathcal{D}_N^\top)^{-1} \mathcal{H}_N - \mathcal{L}_N^\top \bar{\Phi}_{N,t} \mathcal{L}_N \right]^{-1}\\
    \Omega_{\bar{\Phi}_{N,t}} &\coloneqq \Omega_N + \mathcal{J}_N^\top S_{\bar{\Phi}_{N,t}}^{-1} \mathcal{J}_N\\
    \mathcal{J}_N &\coloneqq \mathcal{O}_N^R - \mathcal{L}_N \mathcal{H}_N^\top \left[\mathcal{D}_N\mathcal{D}_N^\top +\mathcal{H}_N \mathcal{H}_N^\top\right]^{-1} \mathcal{O}_N\\
    \Omega_N & \coloneqq \mathcal{O}_N^\top (\mathcal{D}_N \mathcal{D}_N^\top + \mathcal{H}_N \mathcal{H}_N^\top)^{-1}\mathcal{O}_N\\
    S_{\bar{\Phi}_{N,t}} &\coloneqq \mathcal{L}_N (I_{N n_x} + \mathcal{H}_N^\top (\mathcal{D}_N \mathcal{D}_N^\top)^{-1} \mathcal{H}_N)^{-1} \mathcal{L}_N^\top -\bar{\Phi}_{N,t}^{-1} \\
    \mathcal{K}_{\bar{\Phi}_{N,t}} &\coloneqq \begin{bmatrix}
\mathcal{D}_N\mathcal{D}^\top + \mathcal{H}_N \mathcal{H}_N^\top & \mathcal{H}_N \mathcal{L}_N^\top \\ \mathcal{L}_N\mathcal{H}_N^\top & \mathcal{L}_N \mathcal{L}_N^\top - \bar{\Phi}_{N,t}^{-1}
\end{bmatrix}.
\end{split}
\end{equation*}

Importantly, the mapping~\eqref{eqn:downsampled_Riccati} corresponds to an $N$-fold composition of the original DR Riccati mapping $r_\ambset$. Thus, if we can establish that $r_{\ambset,t}^d$ is a contraction for $t \geq q$ with $q$ fixed, then $r_\ambset$ is also a contraction. By the Banach's fixed point theorem~\cite{banach1922operations}, this implies the existence and uniqueness of a fixed point $\Xpriorssnom \in\pd{n_x}$ of $r_\ambset$ satisfying
\[
    \Xpriorssnom = r_\ambset(\Xpriorssnom),
\]
which can be computed by recursively applying~\eqref{eqn:DR_riccati} starting from any $\Xnom\in\pd{n_x}$.

The following proposition establishes sufficient conditions under which the mapping $r_{\ambset,t}^d$ becomes a contraction.

\begin{proposition}\label{prop:phi_tilde}
    Suppose~\Cref{ass:Gauss,ass:time_inv_nom,ass:ctrb_obsv} hold. Let
    \begin{equation}\label{eqn:phi_tilde_cond}
\tilde{\phi}_N = \frac{1}{\lambda_\mathrm{\max}(\mathcal{L}_N (I_{N n_x} + \mathcal{H}_N^\top (\mathcal{D}_N \mathcal{D}_N^\top)^{-1} \mathcal{H}_N)^{-1} \mathcal{L}_N^\top)} > 0.
\end{equation}
Then, there exists $\phi_N \in (0, \tilde{\phi}_N)$ and $N \geq n_x$, such that if, for some fixed $q >0$, the matrix $\bar{\Phi}_{N,t}$ satisfies 
\begin{equation}\label{eqn:phi_cond}
0 \preceq \bar{\Phi}_{N,t} \preceq \phi_N I_{N n_x}, \quad \forall t\geq q, 
\end{equation}
then $r_{\ambset,t}^d$ is a contraction mapping for all $t \geq q$, and consequently $r_\ambset$ is also a contraction mapping.
\end{proposition}

The proof of this proposition can be found in Appendix~\ref{app:phi_tilde}.  A direct approach to find a $\phi_N$ satisfying~\eqref{eqn:phi_cond} is provided in~\cite{zorzi2017convergence}. Specifically, we initially set $\phi_N = \tilde{\phi}N$ and check whether $\Omega{\phi_N I_{Nn_x}}$ is positive definite. If it is not, we iteratively decrease $\phi_N$ until $\Omega_{\phi_N I_{Nn_x}}$ becomes positive semi-definite but singular. Consequently, if $\bar{\Phi}{N,t} \prec \phi_N I{Nn_x}$ for all $t \geq \tilde{q}$, then the Gramians $\Omega_{\bar{\Phi}{N,t}}$ and $W{\bar{\Phi}_{N,t}}$ are positive definite for $k \geq \tilde{q}$. This guarantees that the downsampled DR Riccati recursion in~\eqref{eqn:downsampled_Riccati} is a contraction mapping.

Unfortunately, in practice, condition~\eqref{eqn:phi_cond} is not straightforward to verify. Since $\bar{\Phi}_{N,t}$ depends on the solutions of the SDP problem~\eqref{eqn:DRKF_SDP} at time step $tN+N-1, \dots, tN-1$, it can only be evaluated after solving the sequence of SDPs. Therefore, it becomes necessary to derive tractable sufficient conditions on the ambiguity set radii $\theta_{x}$ and $\theta_v$ that ensure the condition~\eqref{eqn:phi_cond} holds uniformly across all time stages.
To simplify the analysis, we focus on the special case where $\theta_v=0$, indicating that there is no distributional ambiguity in the measurement noise.

\begin{theorem}\label{thm:theta_max}
Suppose~\Cref{ass:Gauss,ass:time_inv_nom,ass:ctrb_obsv} hold and $\theta_v = 0$. Let
\begin{equation}\label{eqn:theta_max}
    \theta_{\max} \coloneqq \sqrt{\frac{\Tr[\bar{\Sigma}_{x,q}^-]}{1 - \phi_N \lambda_{{\max}}(\bar{\Sigma}_{x,q}^-)}} - \sqrt{\Tr[\bar{\Sigma}_{x,q}^-]}
\end{equation}
for a fixed $N \geq n_x$ and $q > 0$, where $\bar{\Sigma}_{x,q}^- \in\pd{n_x}$ is generated according to the standard Riccati recursion
\[
r(\bar{\Sigma}_{x,q}^-):= A((\bar{\Sigma}_{x,q}^-)^{-1} + C^\top \hat{\Sigma}_v^{-1} C)^{-1} A^\top + \hat{\Sigma}_w
\]
with $\bar{\Sigma}_{x,0}^- = \hat{\Sigma}_w$. Then, for any $\theta_x \in [0, \theta_{\max}]$ and any $\Xnom\in\pd{n_x}$, the sequence $\{\Xpriornom\}$ generated by the DR Riccati equation~\eqref{eqn:DR_riccati} converges to a unique fixed point $\Xpriorssnom = r_\ambset(\Xpriorssnom)$. Moreover, as $t\to\infty$, the covariance matrix $\Xprioropt$ obtained as an optimal solution to~\eqref{eqn:DRMMSE_opt} and the corresponding DR Kalman gain matrix $K_t^*$ converge to their steady-state values $\Xpriorssopt$ and
$K_{ss} = \Xpriorssopt C^\top (C \Xpriorssopt C^\top + \hat{\Sigma}_v)^{-1}$, respectively.
\end{theorem}

The proof of this theorem can be found in Appendix~\ref{app:theta_max}.
\Cref{thm:theta_max} provides a systematic approach for determining a conservative yet sufficient Wasserstein ambiguity set radius $\theta_x$ that guarantees convergence of the DR Kalman filter. A useful consequence of~\Cref{thm:theta_max} is that it enables the computation of the steady-state values for the covariance matrix $\Xpriornom$ and the corresponding gain matrix $K_t^*$ via solving a single SDP problem \emph{offline}.

\begin{cor}\label{cor:ss_sdp}
Suppose~\Cref{ass:Gauss,ass:time_inv_nom,ass:ctrb_obsv} hold and $\theta_v = 0$. Let $\theta_x \in [0, \theta_{\max}]$ with $\theta_{\max}$ defined in~\eqref{eqn:theta_max}. Then, the steady-state covariance matrices $(\Xpriorssnom, \Xpriorss, \Xpostss)$ form an optimal solution to the SDP problem~\eqref{eqn:ss_DRKF_SDP}. Consequently, $K_{ss}^* = K_{ss} = \lim_{t\to\infty} K_t^*$.
\end{cor}

An interesting perspective on the DR Kalman filter is obtained by comparing it to other well-known filtering methods. In particular, since the DR Riccati mapping~\eqref{eqn:riccati_like} resembles the risk-sensitive Riccati equation, the time-varying DR Kalman filter can be interpreted as a risk-sensitive filter with risk sensitivity parameter $\theta$ and a time-varying estimation cost weighting matrix $Q_t \in \psd{n_x}$ chosen so that $\theta Q_t^\top Q_t = \Phi_{t-1}$. Similarly, 
 the steady-state DR Kalman filter corresponds to the steady-state risk-sensitive filter (e.g.,~\cite{levy2016contraction}) with parameter $\theta$ and a constant matrix $Q \in \psd{n_x}$ satisfying $\theta Q^\top Q = \Phi_{ss}$, where $\Phi_{ss}$ denotes the limiting value of $\Phi_t$ as $t \to \infty$.
Notably, when $\theta_x = 0$, the DR Kalman filter reduces to the standard Kalman filter applied under the least-favorable measurement noise distribution; specifically, this corresponds to using the covariance matrix $\Sigma_{v,t}^*$ in the time-varying case and $\Sigma_{v,ss}^*$ in the steady-state case.

\section{Simulation Results}\label{sec:experi}

To validate the theoretical findings and evaluate the practical performance of the proposed steady-state DR Kalman filter, we conducted two sets of simulation studies. In the first experiment, we verified the convergence behavior of the DR Kalman filter by comparing the time-varying and time-invariant formulations. The second experiment evaluated the real-world utility of the steady-state DR Kalman filter in a trajectory tracking control task under poorly estimated nominal distributions. Across all experiments, we compared the DR Kalman filter with several baseline filters and demonstrated its superior performance in terms of estimation accuracy and robustness.\footnote{The source code is available online: \url{https://github.com/jangminhyuk/SteadyStateDRKF}}

\subsection{Convergence of DR Kalman filter}

To illustrate the convergence behavior of the DR Kalman filter, we compared the time-varying and time-invariant DR Kalman filters on two systems: a randomly generated system and a benchmark 2D system. In both experiments, the systems were designed to satisfy~\Cref{ass:Gauss,ass:time_inv_nom,ass:ctrb_obsv}. We set $\theta_v = 0$ and computed $\theta_x = \theta_{\max}$ using~\eqref{eqn:theta_max} with $q = 20$. According to~\Cref{thm:theta_max}, the time-varying filter should converge to its steady-state counterpart, which corresponds to the time-invariant DR Kalman filter. 

To compute $\theta_{\max}$, we first evaluated $\tilde{\phi}_N$ using~\eqref{eqn:phi_tilde_cond}, then searched for a suitable $\phi_N$ satisfying the condition $\phi_N \lambda_{\max}(\bar{\Sigma}_{x,q}^-) < 1$ and ensured positive definiteness of the matrix $\Omega_{\bar{\Phi}_{N,t}}$. We initialized $\phi_N = \tilde{\phi}_N$ and iteratively decreased it until $\Omega_{\phi_N I_{Nn_x}}$ became positive semidefinite but non-singular.

\subsubsection{Randomly Generated Systems}
In the first experiment, we randomly generated 100 systems with $n_x = n_y = 2$ and block length $N = 5$. For each system, the matrices $A$ and $C$ were drawn such that $(A, \hat{\Sigma}_w^\frac{1}{2})$ is controllable and $(A, C)$ is observable. The nominal covariance matrices $\hat{\Sigma}_w \in \pd{n_x}$ and $\hat{\Sigma}_v \in \pd{n_y}$ were also randomly sampled. In all cases, the time-varying DR Kalman filter converged successfully to the steady-state solution when using $\theta_x = \theta_{\max}$, as verified by ensuring that the relative trace difference less than 2\%, i.e.,
\[
\left|\frac{\Tr[\Xpostssopt] - \Tr[\Xpost]}{\Tr[\Xpostssopt]}\right|\times 100\% \leq 2\%,
\]
which confirms the theoretical guarantees established in our convergence analysis.

\subsubsection{2D System}

\begin{figure}[t] \centering \includegraphics[width=0.7\textwidth]{convergence_rate_analysis.pdf}
\caption{Logarithmic plot of the relative difference between $\Tr[\Xpost]$ and $\Tr[\Xpostssopt]$ over time.}
\label{fig:convergence} 
\end{figure}

We further evaluated the convergence behavior on a benchmark system from~\cite{zorzi2015convergence} with
\[
A = \begin{bmatrix}
0.1 & 1\\
0 & 1.2
\end{bmatrix}, \quad
C = \begin{bmatrix}
1 & -1
\end{bmatrix}.
\]
The nominal covariance matrices were set to $\hat{\Sigma}_w = I_2$ and $\hat{\Sigma}_v = 1$. We set $\theta_v = 0$ and $\theta_x = \theta_{\max}$, with $\theta_{\max} = 3.356$ computed using~\eqref{eqn:theta_max} for $q=20$ and $N=8$.

To quantify convergence, we tracked the relative trace difference between the posterior covariance matrix $\Xpost$ at each time step and the steady-state posterior covariance matrix $\Xpostssopt$, defined as $|{( \mathrm{Tr}[\Sigma_{x,ss}^{*}] - \mathrm{Tr}[\Sigma_{x,t}])}/{\mathrm{Tr}[\Sigma_{x,ss}^{*}]}|$.
As shown in~\Cref{fig:convergence}, this relative trace difference decays rapidly from approximately $10^0$ to $10^{-5}$ in fewer than 20 iterations. The convergence exhibits an exponential-like decay on a logarithmic scale, indicating the super-linear convergence of the DR Kalman filter.

\begin{table}[ht!]
\centering
\renewcommand{\arraystretch}{1.1} 
\caption{Mean and standard deviation of LQR cost and average MSE under Gaussian and U-Quadratic noise, computed over 20 runs. Best performance for each method is shown in bold.}
\resizebox{\linewidth}{!}{%
\begin{tabular}{
  >{\centering\arraybackslash}p{0.5cm}  
  >{\raggedright\arraybackslash}p{5.2cm}
  *{6}{c}
}
\hline
\multicolumn{2}{c}{Robustness Parameter $\theta$} & 0.1 & 0.2 & 0.4 & 0.5 & 1.0 & 2.0 \\
\hline
\multicolumn{8}{c}{\parbox[c][18pt][c]{\linewidth}{\centering \textbf{Gaussian noise}}}\\
\hline
\multirow{5}{*}{%
  \rotatebox[origin=c]{90}{%
    \begin{minipage}{2.2cm}
    \centering \textbf{LQR Cost}
    \end{minipage}
  }%
} & Time-varying Kalman filter & \multicolumn{6}{c}{723.6 (1051.3)} \\
 & Steady-state Kalman filter & \multicolumn{6}{c}{705.8 (952.8)} \\
  & Risk-sensitive filter & 812.8 (1171.9) & 720.8 (1028.3) & 633.3 (911.4) & \textbf{615.8 (904.5)} & 635.8 (1112.3) & 821.2 (1879.4) \\
 & BCOT filter& 437.9 (100.2) & \textbf{434.5 (120.8)} & 440.8 (154.2) & 443.7 (164.8) & 450.0 (180.4) & 455.0 (180.7) \\
 & Steady-state DR Kalman filter & 137.2 (25.9) & 133.1 (23.1) & \textbf{131.9 (22.1)} & 131.9 (22.0) & 132.4 (21.9) & 132.9 (21.9) \\
\hline
\multirow{5}{*}{%
  \rotatebox[origin=c]{90}{%
    \begin{minipage}{2.2cm}
    \centering \textbf{Average MSE}
    \end{minipage}
  }%
} & Time-varying Kalman filter & \multicolumn{6}{c}{2.690 (4.365)} \\
 & Steady-state Kalman filter & \multicolumn{6}{c}{2.781 (4.397)} \\
  & Risk-sensitive filter & 2.245 (3.694) & 2.003 (3.305) & 1.746 (2.969) & \textbf{1.680 (2.931)} & 1.623 (3.356) & 1.882 (4.971) \\
 & BCOT filter & 1.404 (0.452) & \textbf{1.377 (0.530)} & 1.410 (0.676) & 1.417 (0.706) & 1.447 (0.780) & 1.477 (0.751) \\
 & Steady-state DR Kalman filter & 0.196 (0.043) & 0.188 (0.034) & \textbf{0.189 (0.033)} & 0.191 (0.034) & 0.194 (0.034) & 0.196 (0.034) \\
\hline
\multicolumn{8}{c}{\parbox[c][18pt][c]{\linewidth}{\centering \textbf{U-Quadratic noise}}}\\
\hline
\multirow{5}{*}{%
  \rotatebox[origin=c]{90}{%
    \begin{minipage}{2.2cm}
    \centering \textbf{LQR Cost}
    \end{minipage}
  }%
  } & Time-varying Kalman filter & \multicolumn{6}{c}{245.4 (240.0)} \\
 & Steady-state Kalman filter & \multicolumn{6}{c}{263.9 (304.7)} \\
  & Risk-Sensitive filter & 270.5 (299.6) & 240.5 (254.7) & 206.9 (202.4) & 197.2 (187.2) & \textbf{179.7 (156.5)} & 186.1 (167.7) \\  
 & BCOT filter & 358.3 (37.2) & \textbf{351.2 (33.9)} & 351.8 (38.8) & 360.5 (44.5) & 360.0 (44.3) & 371.7 (48.7) \\  
 & Steady-state DR Kalman filter & 96.7 (9.6) & 95.8 (8.3) & \textbf{95.8 (7.9)} & 95.9 (7.8) & 96.2 (7.7) & 96.3 (7.6) \\
\hline
\multirow{5}{*}{%
  \rotatebox[origin=c]{90}{%
    \begin{minipage}{2.2 cm}
    \centering \textbf{Average MSE}
    \end{minipage}
  }%
 } & Time-varying Kalman filter & \multicolumn{6}{c}{0.856 (1.251)} \\
 & Steady-state Kalman filter & \multicolumn{6}{c}{0.887 (1.415)} \\
  & Risk-Sensitive filter & 0.631 (0.921) & 0.551 (0.802) & 0.449 (0.645) & 0.415 (0.591) & \textbf{0.325 (0.433)} & 0.275 (0.346) \\ 
 & BCOT filter & 1.149 (0.183) & \textbf{1.107 (0.180)} & 1.109 (0.178) & 1.149 (0.194) & 1.148 (0.193) & 1.199 (0.198) \\  
 & Steady-state DR Kalman filter & 0.113 (0.021) & 0.112 (0.018) & \textbf{0.114 (0.017)} & 0.114 (0.017) & 0.116 (0.017) & 0.116 (0.017) \\
\hline
\end{tabular}%
}
\label{tab:combined_results}
\end{table}

\subsection{Performance of the Steady-State DR Kalman Filter}

To evaluate the practical effectiveness of the proposed steady-state DR Kalman filter, we applied it to a 2D trajectory tracking control problem under inaccurate nominal process and measurement noise distributions. The system dynamics are defined as
\[
\begin{split}
x_{t+1} 
&=  \begin{bmatrix} 1 & \Delta t & 0 & 0 \\ 0 & 1 & 0 & 0 \\ 0 & 0 & 1 & \Delta t \\ 0 & 0 & 0 & 1\end{bmatrix} x_t +
\begin{bmatrix} 0.5 \Delta t^2 & 0 \\ \Delta t & 0 \\ 0 & 0.5 \Delta t^2 \\ 0 & \Delta t
\end{bmatrix} u_t + w_t\\
y_t &=  
\begin{bmatrix} 1 & 0 & 0 & 0 \\ 0 & 0 & 1 & 0 \end{bmatrix}  x_t + v_t,
\end{split}
\]
where the time step was set to $\Delta t = 0.2\,\text{s}$. The state vector $x_t = [p_t^x, v_t^x, p_t^y, v_t^y]^\top$ encoded the position and velocity along the $X$ and $Y$ axes, while the control input $u_t = [a_t^x, a_t^y]^\top$ represented acceleration commands.

To demonstrate the importance of DR state estimation under distributional ambiguity, we paired a simple linear-quadratic regulator (LQR) with various filters and compared their performance to our proposed method. The filters evaluated include 
\begin{itemize}
    \item the standard time-varying Kalman filter~\cite{Kalman1960},

    \item the time-invariant (steady-state) Kalman filter~\cite{anderson2005optimal},

    \item the risk-sensitive filter~\cite{whittle1981risk, jacobson2003optimal},  

    \item the bicausal optimal transport-based DR Kalman filter (BCOT filter)~\cite{han2024distributionally}, and 

    \item our steady-state DR Kalman filter.
\end{itemize}
    
The LQR controller was designed to track a smooth curved reference trajectory $x_t^d$ over 50 time steps (i.e., a 10-second interval). The control law was given by $u_t = K_c (x_t^d - \xpost_t)$, where $K_c$ was derived by solving the discrete-time algebraic Riccati equation with weighting matrices $Q = \mathrm{diag}([10,1,10,1])$ and $R = 0.1 I_2$.

We consider two true noise settings:
\begin{enumerate}
    \item Gaussian: $x_0 \sim \mathcal{N}(0, 0.01 I_4)$, $w_t \sim \mathcal{N}(0, 0.01 I_4)$, $v_t \sim \mathcal{N}(0, 0.01 I_2)$,
    \item U-Quadratic: $x_0 \sim \mathcal{UQ}([-0.1, 0.1]^4)$, $w_t \sim \mathcal{UQ}([-0.1, 0.1]^4)$, $v_t \sim \mathcal{UQ}([-0.1, 0.1]^2)$.
\end{enumerate}

\begin{figure}[!ht]
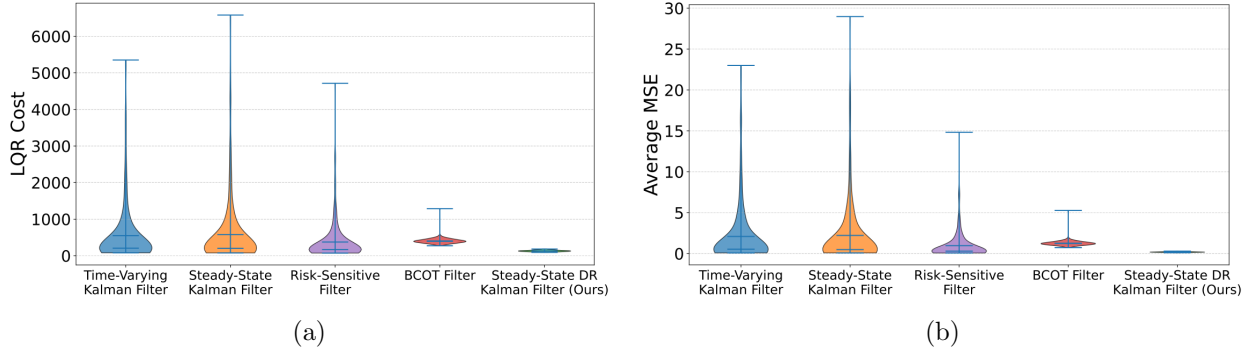

    \centering
    \begin{subfigure}[t]{0.49\textwidth}
        \centering
        \includegraphics[width=\linewidth]{violinplot_lqr_cost_normal.pdf}
        \caption{}
        \label{fig:violin_lqr}
    \end{subfigure}
    \hfill
    \begin{subfigure}[t]{0.49\textwidth}
        \centering
        \includegraphics[width=\linewidth]{violinplot_mse_normal.pdf}
        \caption{}
        \label{fig:violin_mse}
    \end{subfigure}
    \caption{Distributions of LQR cost and average MSE across 200 simulation runs for each filter under Gaussian noise distributions.}
    \label{fig:violin_plots_gaussian}
\end{figure}

\begin{figure}[!ht]
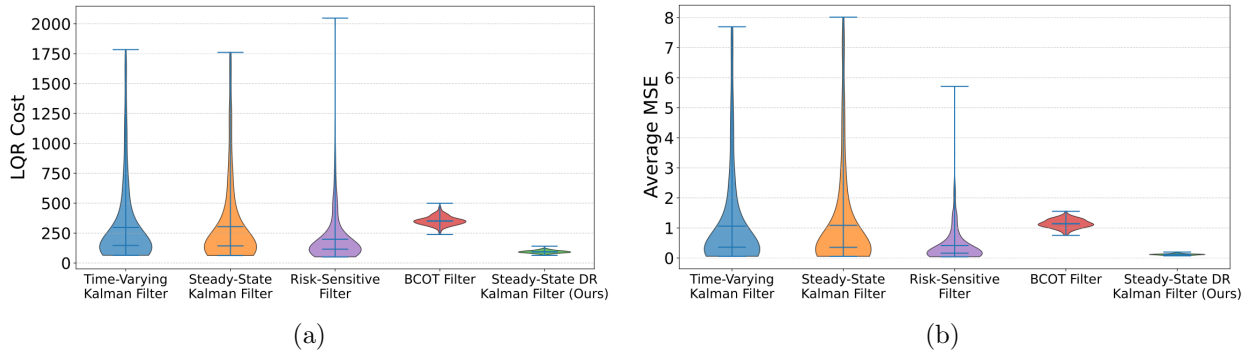

    \centering
    \begin{subfigure}[t]{0.49\textwidth}
        \centering
        \includegraphics[width=\linewidth]{violinplot_lqr_cost_quadratic.pdf}
        \caption{}
        \label{fig:violin_lqr_quadratic}
    \end{subfigure}
    \hfill
    \begin{subfigure}[t]{0.49\textwidth}
        \centering
        \includegraphics[width=\linewidth]{violinplot_mse_quadratic.pdf}
        \caption{}
        \label{fig:violin_mse_quadratic}
    \end{subfigure}
    \caption{Distributions of LQR cost and average MSE across 200 simulation runs for each filter under U-Quadratic noise distributions.}
    \label{fig:violin_plots_uquadratic}
\end{figure}

\begin{figure}[t]
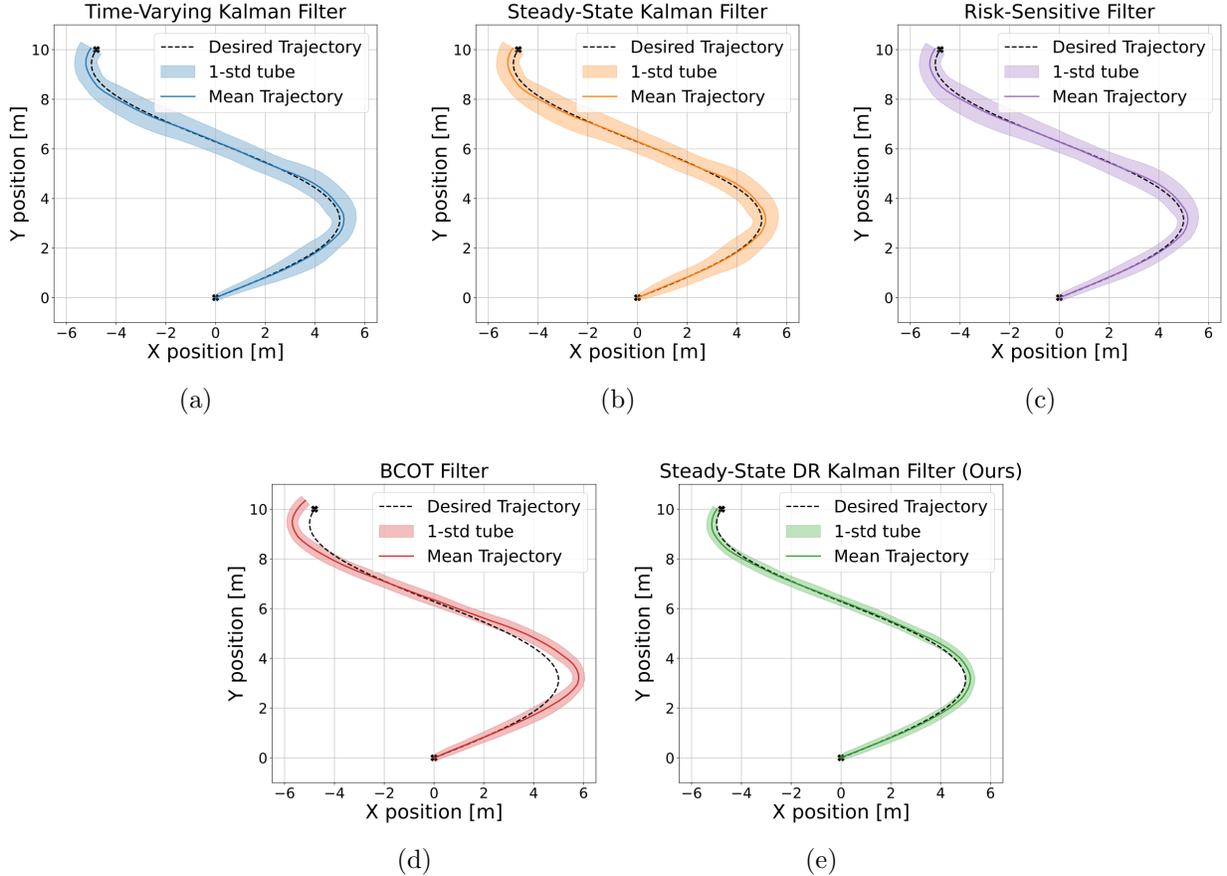

    \centering
    \begin{subfigure}[t]{0.32\textwidth}
        \includegraphics[width=\linewidth]{state_traj_2d_finite_normal.pdf}
        \caption{}
    \end{subfigure}
    \hfill
    \begin{subfigure}[t]{0.32\textwidth}
        \includegraphics[width=\linewidth]{state_traj_2d_inf_normal.pdf}
        \caption{}
    \end{subfigure}
    \hfill
    \begin{subfigure}[t]{0.32\textwidth}
        \includegraphics[width=\linewidth]{state_traj_2d_risk_normal.pdf}
        \caption{}
    \end{subfigure}
    \vskip\baselineskip
    \begin{subfigure}[t]{0.32\textwidth}
        \includegraphics[width=\linewidth]{state_traj_2d_bcot_normal.pdf}
        \caption{}
    \end{subfigure}
    \begin{subfigure}[t]{0.32\textwidth}
        \includegraphics[width=\linewidth]{state_traj_2d_drkf_inf_normal.pdf}
        \caption{}
    \end{subfigure}

    \caption{
        2D tracking performance under Gaussian noise for each filter across 200 runs. 
        The black dashed line shows the desired trajectory, while the colored curve represents the mean trajectory.
        Shaded tubes indicate $\pm$1 standard deviation.
    }
    \label{fig:traj}
\end{figure}

To simulate real-world limitations, the covariance matrices $\hat{\Sigma}_w$ and $\hat{\Sigma}_v$ were estimated from just one second of input-output data using the expectation-maximization method described in~\cite{han2024distributionally}. This setup introduces significant distributional mismatch, underscoring the challenge of learning reliable nominal models from limited data. For both our method and the BCOT  filter, we set $\theta_x = \theta_v = \theta$ to define the ambiguity set radius, while in the risk-sensitive filter, it serves as the risk sensitivity coefficient. For \Cref{fig:traj,fig:violin_plots_gaussian,fig:violin_plots_uquadratic}, we used the best-performing $\theta\in[0.1,2.0]$, i.e., the one yielding the lowest LQR cost.

\Cref{fig:violin_plots_gaussian,fig:violin_plots_uquadratic} illustrate the violin plots showing the distributions of LQR costs and average MSE across all runs. \Cref{tab:combined_results} complements these visualizations by summarizing the mean and standard deviation of both metrics for different values of the robustness parameter $\theta$, with the best-performing $\theta$ highlighted in bold. Across both Gaussian and U-Quadratic noise settings, the proposed filter consistently outperformed all baselines, achieving significantly lower control cost and estimation error. Notably, it maintained reliable performance even under significant distributional mismatches. In contrast, the BCOT filter---being the closest in design to ours---exhibited low variance but a higher mean cost, while the remaining filters displayed considerable performance degradation, as evidenced by long-tailed distributions indicating frequent failures under poorly estimated nominal models.

These results are further confirmed in~\Cref{fig:traj}, which illustrates the average tracking performance across 200 simulation runs under Gaussian noise. Each plot displays the mean estimated trajectory and an associated standard deviation tube for a different filter, using its best-performing $\theta$ value. The time-varying, steady-state, and risk-sensitive Kalman filters exhibit similar behavior, showing reasonably accurate tracking of the desired trajectory on average but with relatively high variance across runs. The BCOT filter, while demonstrating lower variance, tends to deviate from the desired path---particularly in the more curved segments. In contrast, the proposed steady-state DR Kalman filter consistently achieves the most accurate tracking with the narrowest uncertainty tube, highlighting its superior performance and robustness under distributional uncertainties.

\section{Conclusions and Future Work}

In this paper, we proposed a steady-state DR Kalman filter based on Wasserstein ambiguity sets to address state estimation under uncertain noise distributions. Unlike existing time-varying DR filters, our approach computes a steady-state solution by solving a single offline convex SDP problem, thereby enabling efficient implementation. In addition, we derived explicit conditions on the ambiguity radius that guarantee convergence of the time-varying DR filter to its steady-state counterpart. Simulation results confirmed both the theoretical convergence and the superior performance of our method in a tracking control task, outperforming baseline filters. Future work includes developing less conservative convergence conditions and extending the framework to nonlinear systems integrated with DR controllers.

\appendix

\section{Proof of~\Cref{thm:DRKF}}\label{app:DRKF}
\begin{proof}
We prove the theorem by induction. 
From~\Cref{ass:Gauss}, we are given nominal distributions $\hat{\Qdist}_{x,0}^- = \mathcal{N}(\xnom, \hat{\Sigma}_{x,0}^-)$ and $\hat{\Qdist}_{v,0} = \mathcal{N}(\hat{v}_0, \hat{\Sigma}_{v,0})$. By~\Cref{lem:GelbrichMMSE}, the DR-MMSE estimation problem~\eqref{eqn:DRMMSE} admits an equivalent finite-dimensional convex formulation~\eqref{eqn:DRMMSE_opt}, which yields a closed-form DR estimator as in~\eqref{eqn:DR_estimator}. The solution gives the least-favorable prior distribution $\Priordistinitopt = \mathcal{N}(\xnom, \Xpriorinitopt)$ and the least-favorable noise distribution $\Pdist_{v,0}^* = \mathcal{N}(\hat{v}_0, \Sigma_{v,0}^*)$, with $(\Xpriorinitopt, \Sigma_{v,0}^*)$ being the optimal solution of~\eqref{eqn:DRMMSE_opt}. Since both the prior and noise are Gaussian, the resulting posterior distribution is also Gaussian, with mean $\xpost_0 = \phi_0^*(y_0)$ computed from~\eqref{cond_mean} and covariance $\Xpostinit$ from~\eqref{cond_cov}.

Assuming as induction hypothesis that at time $t - 1$ the least-favorable posterior is Gaussian with $\Pdist_{x,t-1}^* = \mathcal{N}(\xpost_{t-1}, \Xpostprev)$, we show the same holds at time $t$. Using the system dynamics and~\Cref{ass:Gauss}, the pseudo-nominal prior distribution at time $t$ is Gaussian, given by $\hat{\Qdist}_{x,t}^- = \mathcal{N}(\xprior_t, \Xpriornom)$, where $\xprior_t$ and $\Xpriornom$ are computed via~\eqref{cond_mean_prior} and~\eqref{cond_cov_prior}. Upon receiving measurement $y_t$, the DR-MMSE estimation problem again admits a Gaussian solution by~\Cref{lem:GelbrichMMSE}. The corresponding least-favorable prior and noise distributions are $\Priordistopt = \mathcal{N}(\xprior_t, \Xprioropt)$ and $\Pdist_{v,t}^* = \mathcal{N}(\hat{v}_t, \Sigma_{v,t}^*)$, with $(\Xprioropt, \Sigma_{v,t}^*)$ solving~\eqref{eqn:DRMMSE_opt}. Consequently, the posterior $\Postdistopt$ is Gaussian with mean $\xpost_t = \phi_t^*(y_t)$ and covariance $\Xpost$ as given in~\eqref{cond_mean} and~\eqref{cond_cov}.
Thus, the DR Kalman filter finds DR state estimates by recursively solving the DR-MMSE estimation problem while preserving the Gaussianity of the least-favorable distributions at every step.
\end{proof}

\section{Proof of~\Cref{prop:phi_tilde}}\label{app:phi_tilde}

\begin{proof}
    The proof follows the argument in~\cite[Proposition 3.1]{zorzi2017convergence}. The downsampled DR Riccati mapping in~\eqref{eqn:downsampled_Riccati} has the same structure as the robust Riccati mapping analyzed in~\cite[Theorem 5.3]{lee2008invariant}. According to this result, if the matrices $\Xpriornomd, \Omega_{\bar{\Phi}_{N,t}}$, and $W_{\bar{\Phi}_{N,t}}$ are positive definite, then $r_{\ambset,t}^d$ is a contraction mapping.

    By \Cref{ass:time_inv_nom}, the matrix $\hat{\Sigma}_w$ is positive definite, which ensures that $\Xpriornomd$ is also positive definite. Moreover, the matrix $\mathcal{Q}_{N,t}$ is positive definite for  $0 \preceq \bar{\Phi}_{N,t} \prec \tilde{\phi}_N I_{N n_x}$, which,  
    under the controllability assumption, is sufficient for the positive definiteness of $W_{\bar{\Phi}_{N,t}}$ for $N \geq n_x$. 
    
    Additionally, the matrix $S_{\bar{\Phi}_{N,t}}$ is negative definite, and the mapping $\bar{\Phi}_{N,t} \mapsto \Omega_{\bar{\Phi}_{N,t}}$ is non-increasing for $0 \prec \bar{\Phi}_{N,t} \prec \tilde{\phi}_N I_{N n_x}$. Under the observability assumption, the matrix $\Omega_0 = \Omega_N$ is positive definite for $N\geq n_x$. Hence, there exists a constant $\phi_N \in (0, \tilde{\phi}_N)$ such that both $W_{\bar{\Phi}_{N,t}}$ and $\Omega_{\bar{\Phi}_{N,t}}$ are positive definite for all $\bar{\Phi}_{N,t}$ satisfying~\eqref{eqn:phi_cond}.

    Therefore, under this condition, the downsampled DR Riccati mapping $r_{\ambset,t}^d$ is a contraction for all $t \geq q$.
\end{proof}

\section{Proof of~\Cref{thm:theta_max}} \label{app:theta_max}
\begin{proof}
Since $\theta_v =0$, we have that $\Sigma_{v,t}^* = \Sigma_{v}$ for all $t$. Then. given a fixed $\phi_N$, our goal becomes identifying the largest admissible value of $\theta_x$, such that
\begin{equation}\label{eqn:phi_cond_1}
\Phi_{t} =  (\Xpriornomnext)^{-1} - (\Xpriornextopt)^{-1} \preceq \phi_N I_{n_x}, \quad t \geq q
\end{equation}
holds for some $q > 0$ fixed. This condition is equivalent to the matrix inequality
\[
\Xpriornextopt \preceq \Xpriornomnext + \phi_N \Xpriornextopt \Xpriornomnext.
\]
Applying the trace operator, we obtain
\[
\Tr[\Xpriornextopt] \leq \Tr[\Xpriornomnext] + \phi_N \Tr[\Xpriornextopt \Xpriornomnext].
\]
Next, invoking von Neumann’s trace inequality yields
\[
\begin{split}
\Tr[\Xpriornextopt \Xpriornomnext] &\leq \sum_i \lambda_i(\Xpriornextopt)\lambda_i(\Xpriornomnext) \\
&\leq \lambda_{{\max}}(\Xpriornomnext) \Tr[\Xpriornextopt],
\end{split}
\]
which leads to the following bound:
\[
\Tr[\Xpriornextopt] \leq \Tr[\Xpriornomnext] + \phi_N \lambda_{{\max}}(\Xpriornomnext) \Tr[\Xpriornextopt].
\]
Thus, if $\phi_N \lambda_{\max}(\Xpriornomnext) < 1$, a sufficient condition to satisfy~\eqref{eqn:phi_cond_1} is:
\begin{equation}\label{eqn:phi_suff}
    \Tr[\Xpriornextopt] \leq \frac{\Tr[\Xpriornomnext]}{1 - \phi_N \lambda_{{\max}}(\Xpriornomnext)}.
\end{equation}
In the case that $\phi_N \lambda_{{\max}}(\Xpriornomnext) \geq 1$, one can always select a smaller $\phi_N$ such that ${\phi}_N \lambda_{{\max}}(\Xpriornomnext) < 1$, while still guaranteeing that both $\Omega_{\bar{\Phi}_{N,t}}$ and $W_{\bar{\Phi}_{N,t}}$ are positive definite.

To connect this to the ambiguity set radius $\theta_x$, we use the following property of the Bures-Wasserstein distance:
\[
\Bures^2(\Xpriornextopt, \Xpriornomnext) \geq \left(\sqrt{\Tr[\Xpriornextopt]} - \sqrt{\Tr[\Xpriornomnext]}\right)^2,
\]
which implies that
\[
\Bures^2(\Xpriornextopt, \Xpriornomnext) \leq \theta_x^2 \quad \implies \quad \Tr[\Xpriornextopt] \leq (\theta_x + \sqrt{\Tr[\Xpriornomnext]})^2.
\]
Combining this with~\eqref{eqn:phi_suff}, we find that choosing $\theta_x \in [0, \theta_{\max}(\Xpriornomnext)]$, with
\[
\theta_{\max}(\Xpriornomnext) = \sqrt{\frac{\Tr[\Xpriornomnext]}{1 - \phi_N \lambda_{{\max}}(\Xpriornomnext)}}- \sqrt{\Tr[\Xpriornomnext]},
\]
will be sufficient for ensuring~\eqref{eqn:phi_cond_1} at any time stage $t$.

Now, from~\cite{levy2016contraction}, it is known that for any $\hat{\Sigma}_{x,0}^- \in \pd{n_x}$, the pseudo-nominal prior covariance matrix satisfies
\[
\Xpriorhat \succeq \bar{\Sigma}_{x,q}^-, \quad  t\geq q+1,
\]
for any $q\geq 0$.
On the other hand, since $\phi_N > 0$, for any $M \in \psd{n_x}$, we have that $0 \leq \phi_N \lambda_{\max}(M) < 1$. Now, suppose $M' \succeq M$. Then, $\Tr[M'] \geq \Tr[M]$ and $\lambda_{\max}(M') \geq \lambda_{\max}(M)$. Combining these, we get that
\[
\sqrt{\frac{\Tr[M']}{1 - \phi_N \lambda_{{\max}}(M')}} \geq \sqrt{\frac{\Tr[M]}{1 - \phi_N \lambda_{{\max}}(M)}}.
\]
Since the first term in $\theta_{\max}(\cdot)$ grows faster than the second term, we conclude that it is monotonically non-decreasing in the L\"owner order.
As a result, we have that
\[
\theta_{\max}(\bar{\Sigma}_{x,q}^-) \leq \theta_{\max}(\Xpriornomnext), \quad t \geq q,
\]
meaning that if we define
$\theta_{\max} := \theta_{\max}(\bar{\Sigma}_{x,q}^-)$ as in~\eqref{eqn:theta_max}, then any $\theta_x \in [0, \theta_{\max}]$ will be sufficient for ensuring~\eqref{eqn:phi_cond_1} for all $t \geq q$.
Hence, for $t \geq \left\lceil \frac{q}{N} \right\rceil$, the downsampled Riccati mapping $r_{\ambset,t}^d$ becomes a contraction. Consequently, the original DR Riccati mapping $r_{\mathcal{D}}$ in~\eqref{eqn:DR_riccati} is also a contraction. By the Banach's fixed-point theorem, it follows that $\Xpriornom$ converges to a unique fixed point
\[
\Xpriorssnom = r_\ambset(\Xpriorssnom).
\]
Thus, the optimal solution $\Xprioropt$ obtained as an optimal solution to~\eqref{eqn:DRMMSE_opt}, or, equivalently, to~\eqref{eqn:DRKF_SDP}, converges to a steady-state solution $\Xpriorssopt$. As a direct consequence, the DR Kalman gain converges to its steady-state value.
\end{proof}

\bibliographystyle{IEEEtran}
\bibliography{ref}

\end{document}